\documentstyle [12pt] {article}

\newcommand{\beq}{\begin{equation}}
\newcommand{\eeq}{\end{equation}}
\newcommand{\beqn}{\begin{eqnarray}}
\newcommand{\eeqn}{\end{eqnarray}}
\newcommand{\eq}[1]{(\ref{#1})}
\newcommand{\bi}[1]{\bibitem{#1}}
\newcommand{\fr}[2]{\frac{#1}{#2}}

\newcommand{\al}{Z\alpha}
\newcommand{\alv}{\mbox{\boldmath $\alpha$ \unboldmath}}
\newcommand{\eps}{\varepsilon}
\newcommand{\g}{\mbox{\boldmath ${\gamma}$\unboldmath}}

\newcommand{\k}{\mbox{${\bf k}$}}
\newcommand{\p}{\mbox{${\bf p}$}}
\newcommand{\lv}{{\mathbf L}}
\newcommand{\r}{{\mathbf r}}
\newcommand{\ro}{\mbox{\boldmath ${\rho}$\unboldmath}}
\newcommand{\ls}{\leq}

\newcommand{\q}{{\bf q}}
\newcommand{\vd}{\mbox{${\bf \Delta}$}}

\begin {document}
\begin{titlepage}
\begin{center}
{\Large \bf Budker Institute of Nuclear Physics}
\end{center}

\vspace{1cm}

\begin{flushright}
{\bf Budker INP 99-76\\
September 15, 1999 }
\end{flushright}

\vspace{1.0cm}

\begin{center}
\Large\bf
On the nature of Coulomb corrections to the
$e^+e^-$ pair production in ultrarelativistic
heavy-ion collisions
\end{center}
\vspace{1.0cm}

\begin{center}
{\bf R.N. Lee, A.I. Milstein}\\
G.I. Budker Institute of Nuclear Physics, \\
630090 Novosibirsk, Russia
\end{center}
\vspace{3cm}

\begin{abstract}
We manifest the origin of the wrong conclusion made by several groups
of authors on the absence of Coulomb corrections to the cross section
of the $e^+e^-$ pair production in ultrarelativistic heavy-ion
collisions. The source of the mistake is connected with an incorrect
passage to the limit in the expression for the cross section. When
this error is eliminated, the Coulomb corrections do not vanish and
agree with the results obtained within the  Weizs\"acker-Williams  approximation.
\end{abstract}

\end{titlepage}

RHIC and LHC projects initiated a set of recent publications on the
$e^+e^-$ pair production in ultrarelativistic heavy-ion collisions.
Using slightly different approaches, the authors of \cite{SW,McL,Gre}
calculated the cross section of the process exactly in the parameters
$\alpha Z_{A,B}$ ($Z_{A,B}$ being the charge numbers of the nuclei $A$
and $B$, $\alpha$ is the fine-structure constant). In these papers
the nuclei were treated as sources of the external field, and the
amplitude was calculated at a fixed impact parameter of the nuclei.
After that the cross section was obtained by the integration over the
impact parameter. As a result, the conclusion was made that
the exact cross section coincides with that calculated in the
lowest order perturbation theory with respect to $\alpha Z_{A,B}$
(Born cross section). On the other hand, in the  Weizs\"acker-Williams  approximation
with respect to one of the nuclei, the cross section of the process
is proportional to the well-known cross section of the $e^+e^-$ pair
production by a photon in a Coulomb field \cite{Bet} and, therefore,
contains the Coulomb corrections. This obvious circumstance was
observed in \cite{Ser}, where the Coulomb corrections in the process
under discussion were calculated.  Though the existence of the
Coulomb corrections is out of doubt, the source of the disagreement
between the results was not revealed so far. This question is
important from the theoretical point of view, since the approach
developed in \cite{SW,McL,Gre} is used now in QCD. In the present
paper we present the solution of this puzzle.

Let the ultrarelativistic nuclei $A$ and $B$ move in the positive and
negative directions of the $z$ axis, respectively. Then the
expression for the cross section of the $e^+e^-$ pair production,
obtained in \cite{SW,McL,Gre}, reads
\begin{eqnarray}
\label{section}
d\sigma&=&\frac{m^2d^3pd^3q}{(2\pi)^6\eps_p\eps_q}
\int
\frac{d^2k}{(2\pi)^2}
|F_B(\k)|^2
|F_A(\q_\perp+\p_\perp-\k)|^2
|{\cal M}(\k)|^2\ ,
\\
{\cal M}(\k)&=&
\overline{u}(p)\frac{\alv
(\k-\p_\perp) +
\gamma_0 m}{-p_+ q_- - (\k-\p_\perp)^2 -m^2 +i\epsilon}
\gamma_-u(-q) +\nonumber
\\
&&+\overline{u}(p)\frac{-\alv
(\k-{\q}_\perp) +
\gamma_0
m}{-p_- q_+ -
(\k-{\q}_\perp)^2 -m^2
+i\epsilon} \gamma_+u(-q)\nonumber\ .
\end{eqnarray}
Here $\p$ and $\eps_p$ ($\q$ and $\eps_q$) are the momentum and
energy of the electron (positron), $u(p)$ and $u(-q)$ are
positive- and negative-energy Dirac spinors,
$\alv=\gamma^0\g$, $\gamma_\pm=\gamma^0\pm\gamma^z$, $\gamma^\mu$
are the Dirac matrices, $p_\pm=\eps_p\pm p^z$, $q_\pm=\eps_q\pm q^z$,
$m$ is the electron mass, $\k$ is a two-dimensional vector lying in
the $xy$ plane, and the function $F(\vd)$ is proportional to the
electron eikonal scattering amplitude in the potential $V(\r)$ of the
corresponding nucleus:
\beq\label{impact}
F(\vd)=\int d^2\rho \exp[-i \ro \vd]
\left\{\exp[-i\chi(\ro)]-1\right\} \ ,\quad
\chi(\ro)=\int\limits_{-\infty}^\infty dz V(z,\ro)\ .
\eeq
For the potential $V(\r)=V_c(r)=-\al/r$, the integral
in $\chi(\ro)$ becomes divergent and requires a
regularization. This regularization can be made by using the
potential $V(\r)=-\al\exp(-r/a)/r$. Performing the integration in
\eq{impact}, and taking the limit $a\to \infty$ at fixed $\vd\not=0$,
one obtains (up to the constant phase depending on $a$):
\beq\label{impact1}
F(\vd)={\cal F}(\vd)\equiv i\pi\al
\frac{\Gamma(1-i\al)}{\Gamma(1+i\al)}
\left(\frac{4}{\Delta^2}\right)^{1-iZ\alpha} \ .
\eeq
Actually, to obtain this result one can use any regularization of the
phase $\chi(\ro)$ for which $\chi(\ro)\to 0$ at $\rho\to\infty$.
Since $|{\cal F}(\vd)|^2=(4\pi\al/\Delta^2)^2\propto Z^2$, then the
substitution \eq{impact1} into \eq{section} would lead to the
wrong conclusion \cite{SW,McL,Gre} that the exact cross section
coincides with the Born result. Let us show that, in order to obtain
the Coulomb corrections in \eq{section}, it is necessary first to
take the integral over $\k$ using the functions $F(\vd)$ with the
regularized phase and then remove the regularization.

Consider the integral
\beq\label{G}
G=\int \fr{d^2k}{(2\pi)^2} \, k^2 \left(
|F(\k)|^2-|F^0(\k)|^2\right) \ ,
\eeq
where $F^0(\vd)=-i\int d\ro\exp(-i\vd\ro) \chi(\ro)$ is the first
term of the expansion of $F(\vd)$ with respect to the potential.
For $F={\cal F}$ and, correspondingly, $F^0={\cal F}^0\equiv
4i\pi\al/\Delta^2$, the integrand in \eq{G} vanishes. Let us show that
the integral $G$ is not equal to zero for the regularized $F$ and is
independent of the regularization method, if $V(\r)\to -\al/r$ at
$r\to 0$ ( when $\chi(\ro)\to 2\al\ln(\rho)+const$ at $\ro\to 0$).
For the sake of simplicity, we present the proof of this statement for
a spherically symmetric potential $V(r)$. Taking the integral in
\eq{impact} over the angle of $\ro$, and integrating by parts
over $\rho$, we obtain the following expression for $F$:
\beq\label{Fbyparts}
F(\k)=\fr{2\pi i}k \int\limits_0^\infty d\rho\,
\rho J_1(k\rho) \chi'(\rho) \exp(-i\chi(\rho))\ ,
\eeq
where $J_1(x)$ is the Bessel function.
The function $F^0(\k)$ can be obtained from \eq{Fbyparts} by omitting
the exponent in the integrand.  Substituting \eq{Fbyparts} into
\eq{G}, and integrating over the angle of $\k$, we find
\beqn\label{G1}
G&=&2\pi\int\limits_0^\infty dq\, q \int\limits_0^\infty\!
\int\limits_0^\infty d\rho_1\, d\rho_2\, \rho_1\rho_2 J_1(k\rho_1)
J_1(k\rho_2)\times\nonumber\\ &&\times\chi'(\rho_1) \chi'(\rho_2)
\left\{\exp[-i\chi(\rho_1)+i\chi(\rho_2)]-1\right\}\ .
\eeqn
If one changes naively the order of integration over $k$ and
$\rho_{1,2}$ and takes the integral over $k$, using the relation
$$
\int\limits_0^\infty dk\, k J_1(k\rho_1)
J_1(k\rho_2) =\fr 1{\sqrt{\rho_1\rho_2}} \delta(\rho_1-\rho_2)\ ,
$$
then, after the integration over $\rho_1$, the result will be zero.
To demonstrate that the change of the integration order in \eq{G1} is
invalid, we restrict the upper limit of the integral over $k$ by
 $Q$. After that one can change  the order of integration in triple
integral in \eq{G1}. Integrating over $k$, we obtain
\beqn\label{G2}
G&=&2\pi
\int\limits_0^\infty\!  \int\limits_0^\infty d\rho_1\, d\rho_2\,
\fr{Q\rho_1\rho_2}{\rho_1^2-\rho_2^2}\left[
\rho_2 J_0(Q\rho_2) J_1(Q\rho_1)-
\rho_1 J_0(Q\rho_1) J_1(Q\rho_2)\right]\times\nonumber\\
&&\times\chi'(\rho_1)
\chi'(\rho_2) \left\{\exp[-i\chi(\rho_1)+i\chi(\rho_2)]-1\right\}\ .
\eeqn
Substituting $\rho_{1,2}\to\rho_{1,2}/Q$, and taking the limit
$Q\to\infty$ with the use of the asymptotics of $\chi$, we find
\beqn\label{G3}
G&=&8\pi(\al)^2 \int\limits_0^\infty\!
\int\limits_0^\infty d\rho_1\, d\rho_2\,
\fr{1}{\rho_1^2-\rho_2^2}\left[
\rho_2 J_0(\rho_2) J_1(\rho_1)-
\rho_1 J_0(\rho_1) J_1(\rho_2)\right]\times\nonumber\\
&&\times
\left\{\left(\fr{\rho_2}{\rho_1}\right)^{2iZ\alpha}-1\right\}\ .
\eeqn
Making the change of variables $\rho_{1,2}=r \exp(\pm t/4)$,
and integrating over $r$, we obtain the non-zero result for the
quantity $G$:
\beqn\label{Gfin}
G&=&8\pi(\al)^2\int\limits_0^\infty dt \fr{\cos(\al t)-1}{\exp(t)-1}
\\
&=&-8\pi(\al)^2[\mbox{Re}\psi(1+i\al)+C]
=-8\pi(\al)^2 f(\al) \nonumber
\ ,
\eeqn
where $C$ is the Euler constant, $\psi(x)=d\ln\Gamma(x)/dx$.
Thus, we come to the remarkable statement: although the main
contribution to the integral in \eq{G} comes from the region of small
$k$, where $|F(\k)|$ differs from $|{\cal F}(\k)|=4\pi\al/k^2$ and
depends on the regularization parameters (the radius of screening),
nevertheless, the integral $G$ itself is a universal function
of $\al$. Note that the integral \eq{G} appears in the theory of
multiple scattering (see  \cite{Mo} where the approximate formula for
this integral was obtained).

Now it is clear, how to derive the Coulomb corrections starting from
the expression \eq{section}. Let us calculate the Coulomb corrections
related to the nucleus $B$ (the contribution of the higher order
perturbation theory with respect to the parameter $Z_B\alpha$). For
this purpose one should replace in \eq{section} the functions
$|F_B|^2$ and $|F_A|^2$ with $|F_B|^2-|F_B^0|^2$ and $|{\cal
F}_A^0|^2$, respectively, keeping the regularization in the functions
$F_B$ and $F_B^0$. The main contribution to the integral is given by
the region of small $\k$. Therefore, we can neglect $\k$ in the
argument of ${\cal F}_A^0$ and expand the matrix element ${\cal M}$
at small $\k$:
\beq \label{ME}
{\cal M}(\k)\approx \k\lv\ ,\  \lv=
\overline{u}(p)\left\{\fr{\alv(\gamma_-/p_+-\gamma_+/q_+)}{(p_-+q_-)}
+\fr{2\gamma_-(\p_\perp /p_+-\q_\perp /q_+)}{(p_-+q_-)^2}\right\}
u(-q)
\ .
\eeq
Using \eq{Gfin} and \eq{ME}, and performing the summation
over electron and positron polarizations, we obtain the following
expression for the Coulomb corrections related to the nucleus $B$:
\beq \label{coul}
d\sigma_B^{c} =\frac {2G_B d^3p\,
d^3q}{(2\pi)^6\eps_p\eps_q}
\fr{|{\cal F}_A^0(\p_\perp+\q_\perp)|^2}
{[p_+q_+(p_-+q_-)]^2}
\left\{
p_+q_+(\p_\perp+\q_\perp)^2-\fr{2(\p_\perp q_+q_-+\q_\perp
p_+p_-)^2}{(p_-+q_-)^2}
\right\}
\ .
\eeq
Here $G_B$ denotes the function $G$ in \eq{Gfin} at $Z=Z_B$.
The Coulomb corrections related to the nucleus $A$ can be
obtained from \eq{coul} by the substitution $Z_A\leftrightarrow Z_B$
and the replacement of indices $-\leftrightarrow +$.

It is necessary to note the following circumstance. Actually, in the
expansion over $Z_A\alpha$ and $Z_B\alpha$ of the differential cross
section $d\sigma/d\p d\q$ in \eq{section} , only the lowest (Born)
term is correct. As for the higher order terms in \eq{section}
(Coulomb corrections), they give the correct result only after
the integration over the directions of the positron (electron)
momentum.  This is due to the fact that the asymptotic form of the
wave functions in \cite{SW,McL,Gre} corresponds to the problem of
scattering, but not to the problem of pair production. If one
calculates the cross section integrated over the direction of $\q$,
then, due to the completeness relation, it is possible to replace the
set of functions containing in asymptotics the converging spherical
wave with the set of functions containing the diverging spherical
wave. Thus, \eq{coul} should be integrated over the angles of $\q$
($\p$). The same trick was made at the recalculation of the
bremsstrahlung cross section integrated over the photon momentum from
the cross section of pair photoproduction integrated over the
positron momentum \cite{Land}. It explains why the Coulomb
corrections \eq{coul} are given by the region of small $\k$, while at
the calculation of the Coulomb correction using the wave functions
with the correct asymptotic behavior the main contribution would come
from the region $k\sim m$. The same situation occurs at the
calculation of bremsstrahlung and pair photoproduction cross
sections, where the Coulomb corrections come from different regions
of momentum transfers.

Let us calculate within the logarithmic accuracy the Coulomb
corrections to the cross section $d\sigma/d\eps_p d\eps_q$ at
$\eps_{p,q}\gg m$. At the integration over the transverse momenta the
main contribution comes from the region
$\Delta=|\p_\perp+\q_\perp|\ll p_\perp,\, q_\perp\sim m$.
The integral over $\Delta$ requires regularization at $\Delta\to 0$.
It is obvious that the lower limit of integration over $\Delta$
coincides with that in the  Weizs\"acker-Williams  method. In the rest frame of the
nucleus $B$ it has the form
$\Delta_{min}=(\eps_p^0+\eps_q^0)/\tilde{\gamma}$, where
$\eps_{p,q}^0$ are the energies of the electron and positron,
$\tilde{\gamma}$ is the Lorentz factor of the nucleus $A$ in this
frame. In the laboratory frame, where the nuclei $A$ and $B$
have the Lorentz factors $\gamma_A$ and $\gamma_B$, respectively, one
has $\Delta_{min}=(p_++q_+)/\gamma_A$.
Using this cutoff, we obtain
\beq\label{WW}
d\sigma_B^{c}=-\fr{4}{\pi
m^2}(Z_A\alpha)^2(Z_B\alpha)^2 f(Z_B\alpha)
\fr{d\eps_p d\eps_q}{(\eps_p+\eps_q)^2}
\left(1-\fr{4\eps_p\eps_q}{3(\eps_p+\eps_q)^2}\right)
\left[\ln \fr{m^2}{\Delta_{1min}^2}+
\ln \fr{m^2}{\Delta_{2min}^2}
\right]\, .
\eeq
The sum of logarithms in this formula corresponds to the
contributions of two kinematic regions: $p^z,\, q^z>0$ and $p^z,\,
q^z<0$. In the first case $\Delta_{1min}=(\eps_p+\eps_q)/\gamma_A$,
and the corresponding term in \eq{WW} is valid at $m\ll \eps_{p,q}\ll
m\gamma_A$. In the second case
$\Delta_{2min}=m^2/(\eps_p+\eps_q)\gamma_A$, and the corresponding
term is valid at $m\ll \eps_{p,q}\ll m\gamma_B$. Performing the
integration over $\eps_{p,q}$ in the regions indicated, one has
\beq\label{WW1}
\sigma_B^{c}=-\fr{28}{9\pi m^2}(Z_A\alpha)^2(Z_B\alpha)^2 f(Z_B\alpha)
\ln^2(\gamma_A\gamma_B) \ .
\eeq
The formulas \eq{WW} and \eq{WW1} can be easily  obtained in the  Weizs\"acker-Williams
approximation using the well-known result for the exact in $Z\alpha$
pair photoproduction cross section in the field of a nucleus. They
coincide with the result of \cite{Ser} (see also \cite{Ku}). It also
follows from the  Weizs\"acker-Williams  method that the contribution of the terms,
containing the higher orders of $Z_A$ and $Z_B$ simultaneously,
can be neglected within our accuracy.

In papers \cite{SW,McL,Gre} the amplitude of $e^+e^-$ pair production
was obtained at fixed impact parameter between the nuclei. Using this
amplitude, it is possible to represent the Coulomb corrections
related to the nucleus $B$ as the integral over the impact
parameter:
\begin{eqnarray}
\label{section1}
d\sigma_B^c&=&\frac{m^2d^3p\,d^3q}{(2\pi)^6\eps_p\eps_q}
\int d^2\rho
\int\!\!\! \int
\frac{d^2k_1}{(2\pi)^2}
\frac{d^2k_2}{(2\pi)^2}
\exp[i(\k_1-\k_2)\ro]{\cal M}(\k_1) {\cal M}^*(\k_2)
\times\\
&&\times
\left[{\cal F}_B(\k_1) {\cal F}_B^*(\k_2)  -
   {\cal F}_B^0(\k_1) {\cal F}_B^{0\,*}(\k_2)
\right]
{\cal F}_A^0(\q_\perp+\p_\perp-\k_1)
{\cal F}_A^{0\,*}(\q_\perp+\p_\perp-\k_2)
\ . \nonumber
\end{eqnarray}
Again, changing the order of integration would lead to zero
result. Indeed, taking the integral over $\ro$ first, we get
the factor $\delta(\k_1-\k_2)$ in the integrand, and, therefore, the
integral over $\k_1$ vanishes due to the relation $|{\cal
F}_B|^2=|{\cal F}_B^0|^2$. Let us demonstrate that, similar to the
case of the integral \eq{G} calculation, the change of the
integration order in \eq{section1} is incorrect, and the result
\eq{coul} also follows from \eq{section1}. For this purpose, we
restrict the region of integration over $\ro$ by the condition
$\rho<R$. After that it is possible to change the order of
integration and take the integral over $\ro$.  Then the main
contribution to the integral over $\k_{1,2}$ comes from the region
$k_{1,2}\ls 1/R$. Since we are going to take the limit $R\to\infty$,
we can replace ${\cal M}(\k_{1,2})$ with $\k_{1,2}\lv$ and neglect
$\k_{1,2}$ in ${\cal F}_A^0(\q_\perp+\p_\perp-\k_{1,2})$.
Then, we have
\begin{eqnarray}
\label{coul1}
d\sigma_B^c&=&\frac{m^2d^3pd^3q}{(2\pi)^6\eps_p\eps_q} |{\cal
F}_A^0(\q_\perp+\p_\perp)|^2 \fr{|\lv|^2}2 \tilde{G}_B\ ,
\\
\tilde{G}_B&=&8\pi(Z_B\alpha)^2
\int\limits_0^\infty\!\!\!\int\limits_0^\infty \fr{dk_1 dk_2}{k_1^2-k_2^2}
\left\{\left(\fr {k_1}{k_2}\right)^{2iZ_B\alpha}-1\right\}
\times\nonumber\\
&&\times
\left[
k_2 R J_0(k_2 R) J_1(k_1 R)-k_1 R J_0(k_1 R) J_1(k_2 R)
\right]
\ . \nonumber
\end{eqnarray}
Comparing the expression for the function $\tilde{G}_B$ with \eq{G3},
we see that $\tilde{G}_B=G_B$. After the summation over the
electron and positron polarizations the formula \eq{coul1} comes into
\eq{coul}. Note that the expression \eq{coul} can be obtained directly
from \eq{section1} by taking the integral over $\k_{1,2}$ in the
region $k_{1,2}<k_0\ll |\p_\perp+\q_\perp|$ and then integrating over
$\rho$ in the infinite limits.

If the impact parameter $\rho$ is restricted by the beam transverse
size $R_0$, then it follows from the above consideration that
the effect of the finite size appears when we can not neglect $k\sim
1/R_0$ in the argument of ${\cal F}_A^0$ in comparison with
$|\q_\perp+\p_\perp|$. This is equivalent to the condition $R_0\ll
1/\Delta_{min}\sim \gamma_A\gamma_B/m$.

Thus, the method developed in \cite{SW,McL,Gre} can be used
for the calculation of the Coulomb corrections to the $e^+e^-$ pair
production cross section integrated over the direction of the
positron (electron) momentum. Its careful application leads to the
correct result.

We are grateful to V.M.~Katkov and V.M.~Strakhovenko for useful
discussions.

\newpage

\begin {thebibliography} {99}

\bi {SW}
B. Segev, J.C. Wells, Phys. Rev. A {\bf 57}, 1849 (1998);
physics/9805013.

\bi{McL}
A.J. Baltz, L. McLerran, Phys. Rev. C {\bf 58}, 1679 (1998).

\bi{Gre}
U. Eichmann, J. Reinhardt, S. Schramm, and W. Greiner,
Phys.Rev. A {\bf 59}, 1223 (1999);

\bi{Bet}
H. Bethe, L.C. Maximon, Phys. Rev. {\bf 93}, 768 (1954);
H. Davies, H. Bethe, L.C. Maximon, Phys. Rev.
{\bf 93}, 788 (1954).

\bi{Ser}
D.Yu. Ivanov, A. Schiller, V.G. Serbo, Phys. Lett. B
{\bf 454}, 155 (1999)

\bi {Mo}
G. Moli\`ere, Z. Naturforsch. {\bf 2a}, 133 (1947).

\bi {Land}
V.B. Berestetskii, E.M. Lifshitz, L.B. Pitaevskii,
Quantum Electrodynamics, \par 98 (Nauka, Moscow, 1989).

\bi{Ku}
D.Yu. Ivanov, E.A. Kuraev, A. Schiller, V.G. Serbo,
Phys. Lett. B
{\bf 442}, 453 (1998).

\end {thebibliography}

\end {document}